# Isotopic Production Cross Sections in Proton-$^{12}$C Interactions for Energies from 10 MeV/N to 100 GeV/N

Francis A. Cucinotta[*,1], Sungmin Pak[1]

University of Nevada Las Vegas, Las Vegas NV 89154, USA

*Correspondence author
E-mail: francis.cucinotta@unlv.edu

Keywords: Nuclear interaction cross sections, target fragmentation, space radiation, carbon cancer therapy, and proton cancer therapy

**Abstract:**

Proton interactions with $^{12}$C nuclei are a frequent nuclear interaction leading to secondary radiation in tissues for space radiation and cancer therapy with protons or $^{12}$C beams. The fragmentation of $^{12}$C by protons produces a large number of heavy ion (A>4) target or projectile fragments often with high ionization density. Here we develop an analytical model of energy dependent proton-$^{12}$C cross sections for isotopic nuclei production. Using experimental data and a 2$^{nd}$ order optical model an accurate formula for the p-$^{12}$C absorption cross section from <10 MeV/n to >10 GeV/N is obtained. The energy dependence of the elemental and isotopic cross sections is modeled as multiplicities scaled to absorption cross section with average isotopic fractions estimated from experimental data. We show that this approach results in an accurate analytic formula model over the full energy range in Hadron therapy and space radiation protection studies.

1. **Introduction**

High energy carbon ions are used in radiation cancer therapy [1,2] and occur in space radiation exposures to astronauts [3,4]. In space radiation exposures, proton interactions with $^{12}$C are a frequent nuclear interaction due to the $^{12}$C constituents in tissue atoms, and use of polyethylene, or other possible materials (graphite, plastics, etc.) used for equipment of shielding within spacecraft. In the GCR heavy ion spectrum, $^{12}$C has a large abundance amongst heavy ion species, with an energy spectrum that extends from low energies (<10 MeV/N) to very high energies (>10 GeV/N). An important contributor to the biological effects of space radiation or Hadron cancer therapy are the high linear energy transfer (LET) heavy ions produced in the fragmentation of nuclei. Therefore, it is important for radiation transport codes to include accurate models of proton induced fragmentation cross sections. We recently developed a model for p-$^{16}$O isotopic production cross sections [5], while here a similar model is developed for p-$^{12}$C interactions.

In this paper we describe a model of energy dependent nuclear interaction cross sections for the heavy ion (HI) fragments (A>4) produced in p-$^{12}$C nuclear interactions. Theoretical descriptions of p-$^{12}$C (reviewed in [6,7]) include compound nucleus formation and decay at lower energies, knockout and inter-nuclear cascade, and the two-step abrasion-ablation models at higher energies. However, our approach is to develop analytic formulae that accurately describe experimental data for energy dependent nuclear absorption cross sections, elemental and isotopic production cross sections. We introduce a model of elemental production multiplicities scaled to the absorption cross sections and show that the energy dependence for the Z=5 and 6 fragments decrease at higher energies (>1 GeV/N) relative to the absorption cross section, while the energy dependence of the Z= 3 and 4 fragments present modest increases compared to energy dependence of the absorption cross section alone above 100 MeV/N. Our analytic formulae can improve radiation transport code computational speed, while accurately describing the energy dependent HI fragment production cross sections in p-$^{12}$C interactions.

2. **Absorption and Fragmentation Cross Section Model**

We considered experimental data for absorption, and elemental and isotopic production cross sections [8-27] to develop a model of isotopic fragmentation cross sections. Previously we developed a 2nd-order optical model evaluated in the Eikonal approximation to make *ab-initio* predictions of absorption cross sections [28-30]. A first-order model provides predictions by using a double folding of the projectile and target ground-state densities with the free nucleon-nucleon (NN) cross section, while the 2$^{nd}$-order optical model includes a description of two-body nucleon correlations in the nuclear ground state wave function. We include low energy corrections for Coulomb scattering and a medium modified nucleon-nucleon cross section [31] with values used for the free NN parameter described in [32]. **Figure 1** shows the results of a 2$^{nd}$ order optical model calculation with low energy Coulomb and medium modified nucleon-nucleon cross section. The agreement of the 2$^{nd}$ order optical model with absorption experiments is excellent above 300 MeV/A, however shows some differences at lower energies. This is likely due to the simplicity of the treatment nuclear medium effects and need to consider corrections to the Eikonal approximation, such as using solutions to a one-body optical model using the Lippman-Schwinger equation, which avoids the Eikonal approximation [32]. Thus, for our data-base we fit a function in a piece-wise manner with the higher energies (~320 MeV/N) fit directly to the 2$^{nd}$ order optical model and lower energies fit to the model adjusted to experiments as given by (all cross sections are in units of mb):

$$\sigma_{abs}(T) = \begin{cases} \left\{ \dfrac{167701 \exp[-(\log(T/60.57)/0.931)^2]}{T} \right\}, T \leq 48.7 \, MeV/N \\ \max(166.9 + 8120.5/T, 212), 48.7 < T < 323 \, MeV/N \\ = 245 + \dfrac{5.864 \times 10^4}{T} - \left(\dfrac{7430.3}{T}\right)^2 + \left(\dfrac{2197}{T}\right)^3, T \geq 323 \, MeV/N \end{cases} \quad (1)$$

The lower energy term (E<48.7 MeV/N) is based on several experimental results that show significant variations at similar energies (**Figure 1**). Here a regression fit showed P<0.0001 for each parameter in the low energy part of Equation (1) (16701±922, 0.93±0.05, and 60.57±6.61), with an adjusted R$^2$ value of 0.6265. A subjective method to choose optimal data based on the various measurement techniques employed would likely improve the adjusted R$^2$ value, however was not considered.

We parameterized available data for elemental fragment production cross sections using a multiplicity scaled to the energy dependent absorption cross sections for $T_{lab}$>50 MeV/N:

$$\sigma_F(Z_F, T_{lab}) = \sigma_{abs}(T_{lab})[A_{Z_F} + B_{Z_F} \ln(T_{lab})] \quad (2)$$

The second term on the right-hand side of Eq. (2) describes the deviation of the energy dependence of the multiplicity from the energy dependence of the absorption cross section alone. Results for the elemental production multiplicities are given in **Figure 2,** which shows results of a linear regression model fit to the data with results provided in **Table 1**. For Z= 5 and 6 fragments there is a decrease in the production multiplicities with increasing energy above a few hundred MeV/N, while the Z=3 and 4 fragment multiplicities increase with energy. This is likely due to the increased centrality of the collisions with a concomitant increase in the excitation energy of pre-fragment in the ablation stage of fragment production. This decrease will lead to increases in light ion (Z≤2) fragments, however are not considered in the present report.

At lower energies (<50 MeV) some corrections are needed. Most importantly at sufficiently low of an energy not all the fragment production channels are open, and here the threshold energies for various fragment production channels need to be described [10]. Also, pickup and stripping reactions should be considered however such reactions are not in the scope of the present work. We introduced a threshold energy correction to the low energy cross sections of the form:

$$\sigma_F(A_F, Z_F, T_{lab}) = \sigma_F(Z_F, T_{lab}) F(A_F, Z_F)(1 - \exp(-(T_{lab} - E_{th})/E_s))^4 \qquad (2)$$

where $F(A_F, Z_F)$ is value from **Table 2**, $E_{th}$ is the threshold energy for producing the fragment [10], and $E_s$=15 MeV for most fragments. The height of the low energy peak is sensitive to the value of the absorption cross section and $E_s$ with values from 10 to 20 MeV found when data was available. In the absence of any low energy (<50 MeV) data for a specific fragment we chose 15 MeV.

For Z=6 there is detailed experimental data on the production cross sections for $^{11}$C and $^{10}$C production at lower energies that we considered [23,24,26]. **Figure 3** shows experimental results for $^{11}$C versus proton kinetic energy and a fit equation given by:

$$\sigma_{^{11}C}(T) = \sigma_{abs}(T)(A + B\ln(T))[1 - e^{-(T-10)/20}]^{4.1}[1 + 1.6e^{-T/30}]^{2.5} \qquad (3)$$

With $A$ = 0.44 and $B$=-0.047. These values differ modestly compared to the values for Z=6 found in **Table 1**, which is based only on high energy data (>100 MeV/n). For the energy dependence parameter, $B$, because there is sparse information at the higher energies, we set the multiplicity factor to its value at 1 GeV/N for E> 1 GeV/N.

We also studied the isotopic dependence of fragment production, which showed very little energy dependence across several data sets [15-25]. However as shown in **Figure 4** for $^7$Be there is an energy dependent increase with decreasing energy below 200 MeV/N. We then considered data for $^7$Be production at lower energies (**Figure 5**). The data compiled by Read and Viola [10] shown by the red triangles show significant variation at lower energies, tend to predict higher values compared to recent data at energies of a 200-1000 MeV/N, and agree with the more recent measurement above 1 GeV/N. We summarize these observations using the following:

$$\sigma_{^7Be}(T) = \sigma_{abs}(T)(A + B\ln(T))[1 - e^{-(T-23)/20}]^4[1 + 7e^{-T/24}]^{3.2} \qquad (4)$$

Equation (4) favors the data at high energies from more recent experiments that used various types of spectrometers or bubble chambers over older measurements where cross section estimates are based on gamma-ray spectroscopy.

We averaged results for isotopic production cross sections from experiments at several energies, with **Table 2** listing the average multiplicity isotopic fractions for each fragment charge group that are used in our cross-section data base. This approach excludes the $^{11}C$ and $^7Be$ cross sections where the fits of equation (2) and (3) are used. Results of the model for several isotopic production cross sections are described by **Figure 6**, which show a good representation of the available cross sections over a wide energy range. The values for beryllium isotopes could be improved through the use of an energy dependent isotopic fraction as noted above. The nearly constant cross section values for most fragments for $T_{lab}$> 1 GeV/N are consistent with the behavior of the overall absorption cross section.

3. **Discussion and Conclusions**

Fast computational models for nuclear interaction cross sections are advantageous for transport code applications in space research and Hadron therapy investigations. The BRYNTRN and HZETRN codes developed by Wilson et al. [33-35] use data files and parameterizations to achieve fast computational speeds in transport solutions. The present approach allows for very accurate representation of the energy dependent isotopic heavy ion cross sections from p-$^{12}C$ nuclear interactions, which is a frequent interaction in space radiation problems or cancer therapy with carbon or proton beams. Our approach integrates a large number of experimental studies; however, we note that the different experimental methods used could introduce systematic differences that are difficult to quantify. This includes more recent data compared to data available in models developed to study the origin and composition of cosmic rays [36,37], which typically ignore cross section variations below 100 MeV/N, which are of importance for Hadron therapy applications.

The present model goes beyond some simplifications used in the HZETRN/BRYNTRN codes, by considering threshold energy effects on fragment production and providing a more detailed energy dependent model. HZETRN/BRYNTRN data bases considers only threshold dependence for the total absorption cross sections, which introduces errors at lower energies (<50 MeV/N) where individual fragments will have differential thresholds. This deficiency is removed by using the model developed in this paper. This also makes the code data-bases more applicable to Hadron therapy applications where beam energies are generally lower than cosmic rays, while maintaining advantageous CPU efficiency. In future work, a fast-analytic model of double different cross sections for heavy ion fragments will be presented, and information on the $^4$He and other light ion secondaries will be included in the models to extend the approach to lower mass fragments. For heavy ion reactions we have developed abrasion-ablation models [38-40] to form data-files of interaction cross sections that are readily implemented into transport codes. The present approach will be considered to amend these models to improve their accuracy, especially at lower energies.

## Acknowledgements

We acknowledge support from the University of Nevada Las Vegas and the National Cancer Institute (RO1CA208526-01).

**Table 1.** Linear regression fit to energy dependent multiplicities for elemental fragment production after scaling to absorption cross section. Model parameters for $\sigma(Z_F,T)=[A+B\ln(T)]\sigma_{abs}(T)$ where T is the laboratory frame kinetic energy in units of MeV/N.

| Fragment charge group, $Z_F$ | A | B, MeV$^{-1}$ |
|---|---|---|
| 3 | 0.0552±0.0197 | 0.0075±0.0027 |
| 4 | 0.048±0.0106 | 0.0028±0.0015 |
| 5 | 0.3704±0.04 | -0.0266±0.0057 |
| 6 | 0.42±0.04 | -0.0406±0.0061 |

**Table 2.** Energy independent isotopic fractions for elemental groups with Z= 3 to 6 that are scaled to elemental production cross section.

| Fragment Charge Number, $Z_F$ | Fragment Mass Number, $A_F$ | Isotope fraction |
|---|---|---|
| 6 | 11 | 0.92 |
| 6 | 10 | 0.073 |
| 6 | 9 | 0.007 |
| 5 | 12 | 0.002 |
| 5 | 11 | 0.651 |
| 5 | 10 | 0.337 |
| 5 | 8 | 0.01 |
| 4 | 10 | 0.18 |
| 4 | 9 | 0.309 |
| 4 | 7 | 0.511 |
| 3 | 9 | 0.035 |
| 3 | 8 | 0.015 |
| 3 | 7 | 0.42 |
| 3 | 6 | 0.53 |

**Figure 1**. Energy dependent absorption cross sections for protons interactions with $^{12}C$. Experimental data is from references [8-14]. The solid line shows predictions from the 2nd order optical model with medium and Coulomb cross sections, while the dashed line is our preferred fits described by equation (1).

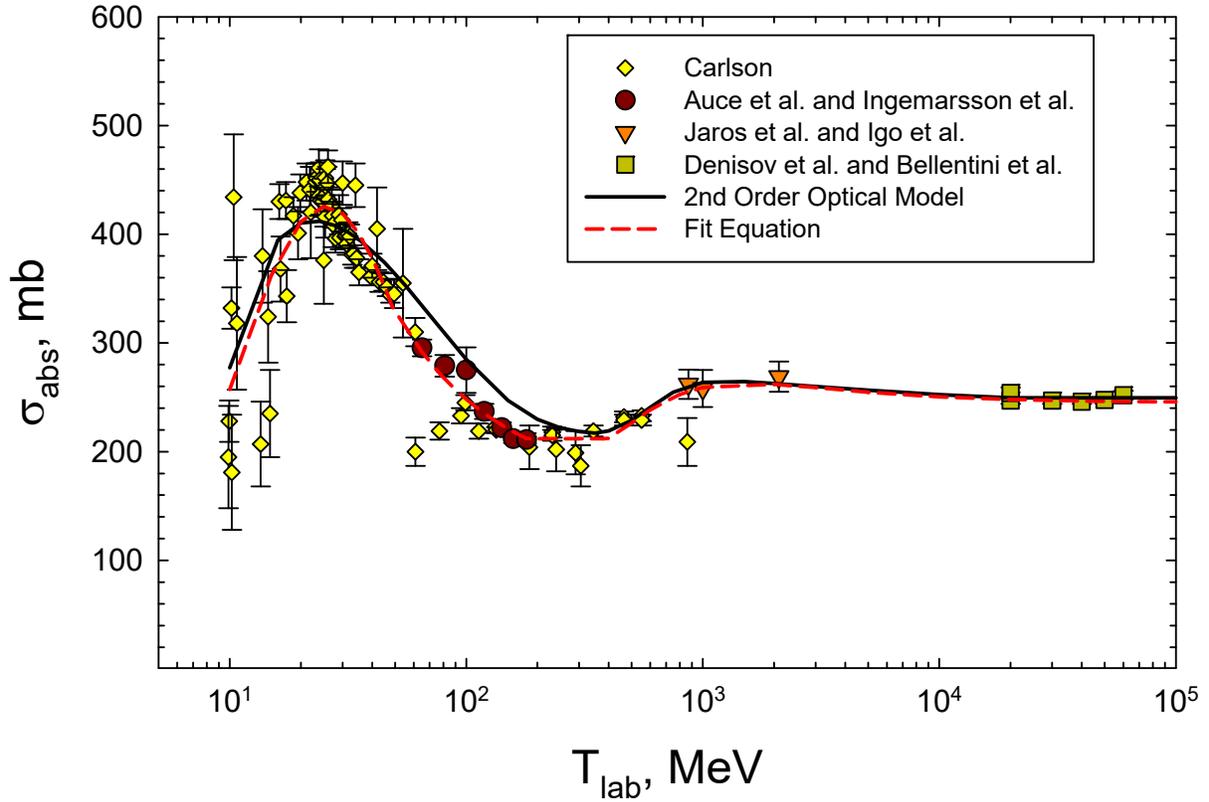

**Figure 2.** Multiplicity for elemental production cross section after scaling to the p-$^{12}$C absorption cross section. Experimental data are from references [15-25]. Results of linear regression model are also shown with parameter values listed in Table 1.

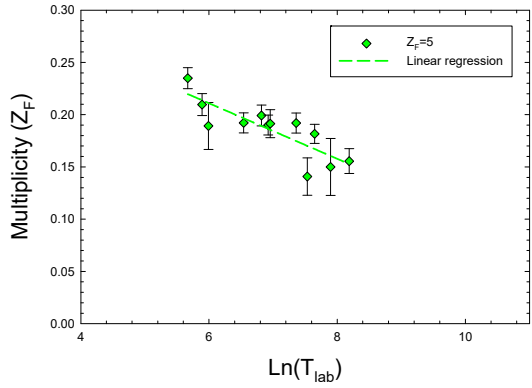

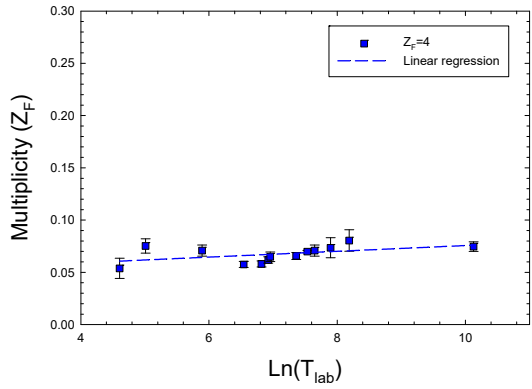

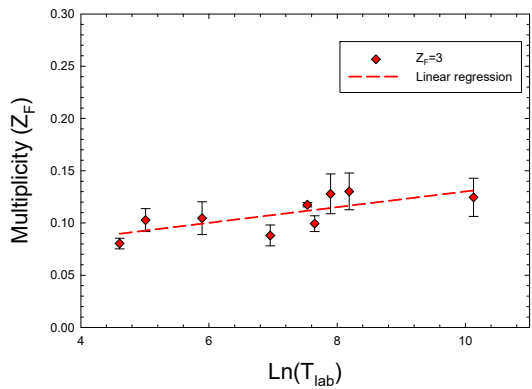

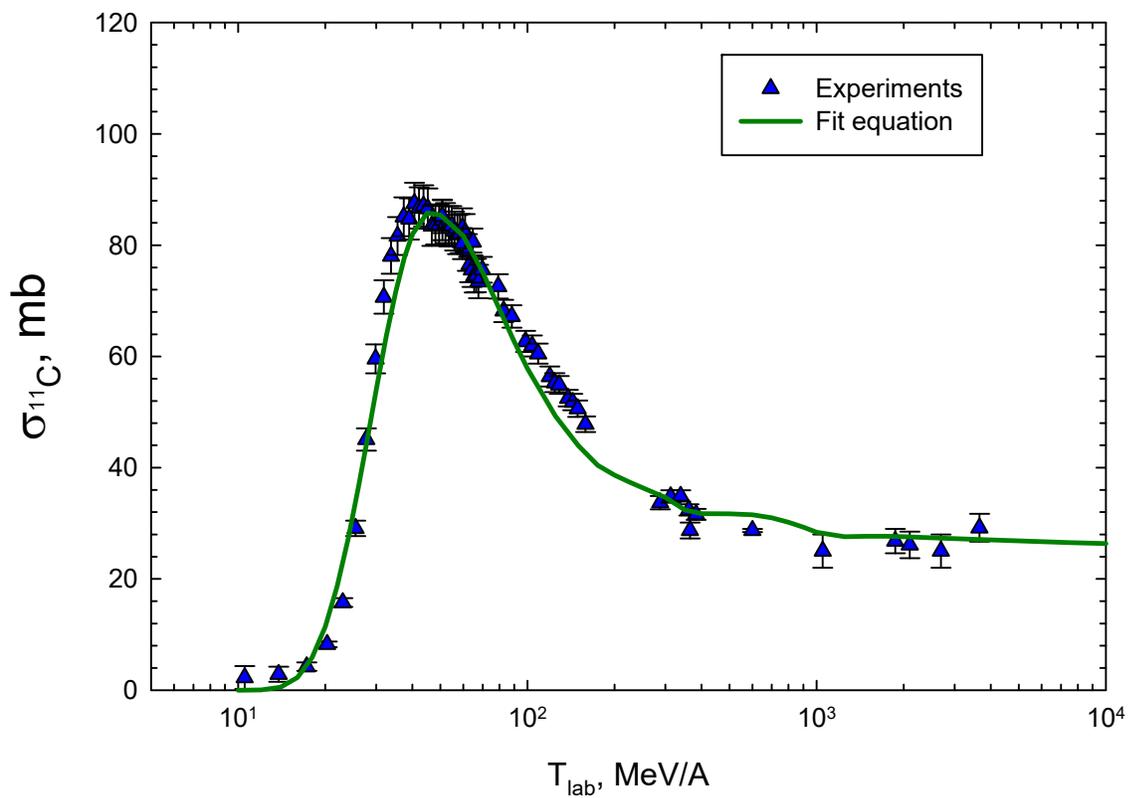

**Figure 3.** Experimental results [15,18,19,21,21-25] for $^{11}$C production in p-$^{12}$C interactions versus laboratory kinetic energy.

**Figure 4.** Isotopic fractions versus lab frame kinetic energy. Results suggest largely constant fractions at higher energies, however with possible energy dependent values at lower energies (<200 MeV/N).

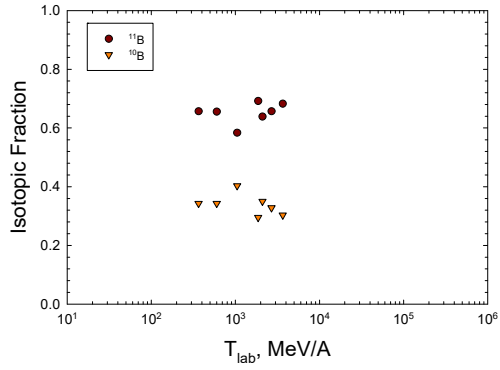

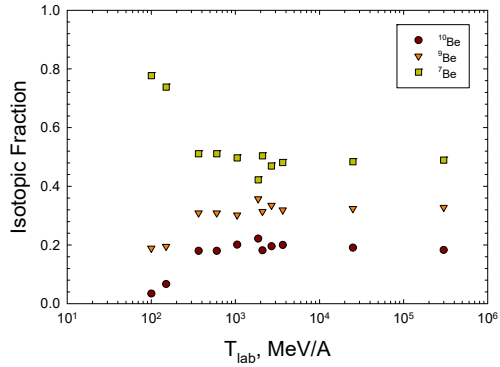

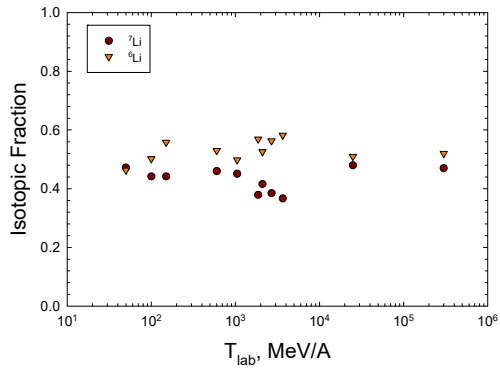

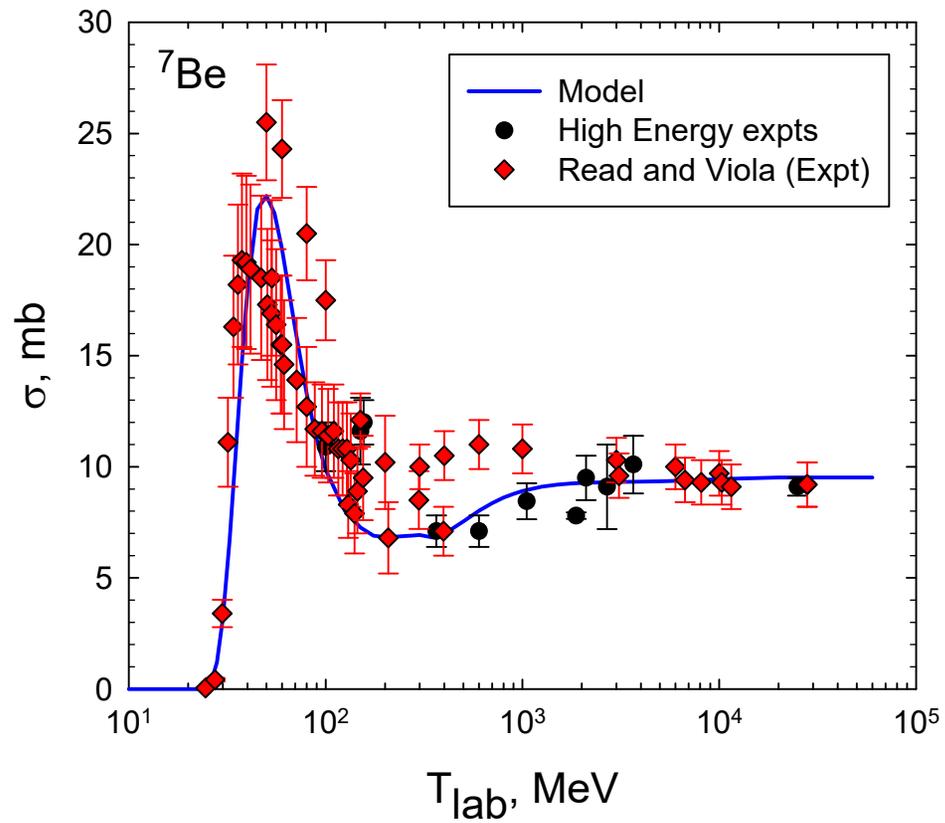

**Figure 5.** Comparison of model to experiment for production of $^7$Be in proton-$^{12}$C interactions ($^{12}$C(p,X)$^7$Be) versus kinetic energy in the laboratory frame of reference. The red triangles are date from the Read and Viola compilation [10]. We have applied a ±20% standard deviation (SD) when SD are not listed by Read and Viola, based on discussion from Williams and Fulmer [27].

**Figure 5.** Comparison of the model for isotope production cross sections for several fragments in p-$^{12}$C interactions to experimental data [15-26]. Results for $^{11}$B, $^{10}$C, $^{10}$B, $^{9}$Be, $^{7}$Li and $^{6}$Li are shown.

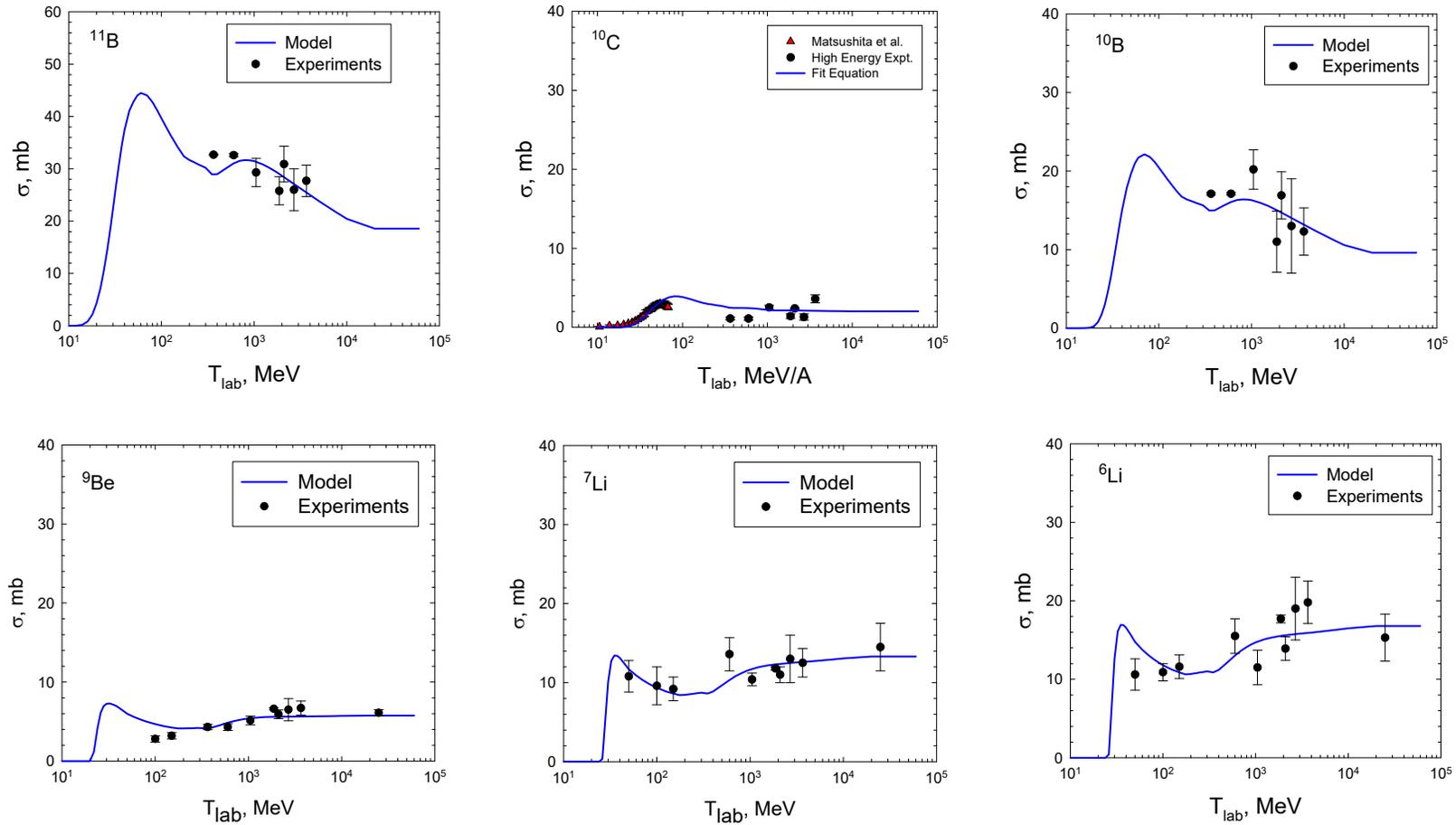